\documentclass[prd,showpacs,nofootinbib,showkeys]{revtex4}

\usepackage{graphicx}
\usepackage{amsmath}
\usepackage{amssymb}
\usepackage{amsxtra}

\begin{document}

\title{General Majorana Neutrino Mass Matrix from a Low Energy $SU(3)$ Family Symmetry with Sterile Neutrinos}

\author{Albino Hern\'andez-Galeana}

\email{albino@esfm.ipn.mx}

\affiliation{ Departamento de F\'{\i}sica,   Escuela Superior de
F\'{\i}sica y Matem\'aticas, I.P.N., \\
U. P. "Adolfo L\'opez Mateos". C. P. 07738, M\'exico, D.F., M\'exico. }



\begin{abstract}
Within the framework of a local $SU(3)$ family symmetry model, we report a general
analysis of the mechanism for neutrino mass generation and mixing, including light sterile
neutrinos. In this scenario, ordinary heavy fermions, top and bottom quarks and tau
lepton, become massive at tree level from Dirac See-saw mechanisms
implemented by the introduction of a new set of $SU(2)_L$ weak singlet vector-like
fermions, $U,D,E,N$, with $N$ a sterile neutrino. Right-handed and the
$N_{L,R}$ sterile neutrinos allow the implementation of a $8\times 8$ general Majorana neutrino mass
matrix with four or five massless neutrinos at tree level. Hence, light fermions, including
light neutrinos get masses from radiative corrections mediated
by the massive $SU(3)$ gauge bosons. We report the corresponding Majorana
neutrino mass matrix up to one loop. Previous numerical analysis of the free parameters show out
solutions for quarks and charged lepton masses within a parameter space region where the vector-like
fermion masses $M_U\,,\,M_D\,,\,M_E$, and the $SU(3)$ family gauge boson masses lie in the low
energy region of $\mathcal{O} (1-20)\,$TeV, with light neutrinos within the correct order of
square neutrino mass differences: $m_2^2-m_1^2 \approx 7 \times 10^{-5}\;\text{eV}^2$,
$m_3^2-m_1^2 \approx 2 \times 10^{-3}\;\text{eV}^2$, and at least one sterile neutrino of the order
$\approx 0.5\;\text{eV}$. A more precise fit of the parameters is still needed to account also for
the quark and lepton mixing.
 \end{abstract}

\maketitle

\tableofcontents


\section{ Introduction }

Although the standard picture with three light flavor neutrinos has been successful to describe
the neutrino oscillation data. On the other hand, there have been recent hints
from the LSND and MiniBooNe short-baseline neutrino oscillation experiments\cite{MiniBooNE,LSND-MiniBooNe}
on the possible existence of at least one light sterile neutrino in the eV scale, which mix with the active neutrinos. On the other hand, an explanation of the strong hierarchy of quark and charged lepton masses
is still a big challenge in particle physics. This hierarchy have suggested to many
model building theorists that light fermion masses could be generated from
radiative corrections, while those of the top and
bottom quarks   and  the tau lepton are generated at
tree level. This may be understood as the breaking of a symmetry among families ,
a horizontal symmetry.

In this report we update the general features of a "Beyond the Standard Model"(BSM) proposal
which introduces a $SU(3)\;$\cite{albinosu32004} gauged family symmetry\footnote{See \cite{albinosu32004,albinosu3bled} and references therein for some $SU(3)$ family symmetry models.}
commuting with the Standard Model group. Previous reports\cite{albinosu3bled}
within this scenario showed that this model has the
features and particle content to account for a realistic spectrum of charged fermion
masses and quark mixing. This BSM model introduce a hierarchical mass
generation mechanism in which the light fermions obtain masses
through one loop radiative corrections, mediated by the massive
bosons associated to the $SU(3)$ family symmetry that is
spontaneously broken, while the masses for the top and bottom
quarks as well as for the tau lepton, are generated at tree level
from "Dirac See-saw"\cite{SU3MKhlopov} mechanisms implemented by
the introduction of a new generation of $SU(2)_L$ weak singlets
vector-like fermions.

\vspace{3mm}
\emph{The $SU(3)$ family symmetry model allows one to address the problem of quark
and lepton masses and mixing, including active and light sterile neutrinos.}

\section{$SU(3)$ flavor symmetry model}

\subsection{Fermion content}

Before "Electroweak Symmetry Breaking"(EWSB) all ordinary, "Standard Model"(SM) fermions
remain massless, and the global symmetry in this limit of all quarks and leptons massless, including
R-handed neutrinos, is:

\begin{eqnarray}
SU(3)_{q_L}\otimes SU(3)_{u_R}\otimes SU(3)_{d_R}\otimes
SU(3)_{l_L}\otimes SU(3)_{\nu_R}\otimes SU(3)_{e_R} \label{globalsymmetry} \\ \nonumber\\
\supset SU(3)_{q_L+u_R+d_R+l_L+e_R+\nu_R} \equiv SU(3) \label{su3symmetry}
\end{eqnarray}

\vspace{2mm}

\noindent We define the gauge group symmetry $G\equiv SU(3) \otimes G_{SM}$
, where Eq.(\ref{su3symmetry}) defines the $SU(3)$ gauged family symmetry,
and $G_{SM}\equiv SU(3)_C \otimes SU(2)_L \otimes U(1)_Y$ is the
"Standard Model" gauge group, with $g_H$, $g_s$, $g$ and $g^\prime$ the corresponding
coupling constants. The content of fermions assumes the ordinary quarks and
leptons assigned under G as:

\begin{equation*}
\psi_q^o = ( 3 , 3 , 2 , \frac{1}{3} )_L  \qquad ,\qquad \psi_u^o = ( 3 , 3, 1 , \frac{4}{3} )_R
\qquad ,\qquad \psi_d^o = (3, 3 , 1 , -\frac{2}{3} )_R  \end{equation*}

\begin{equation*} \psi_l^o =( 3 , 1 , 2 , -1 )_L \qquad , \qquad \psi_e^o = (3, 1 , 1,-2)_R  \;, \end{equation*}

\noindent where the last entry corresponds to the
hypercharge $Y$, and the electric charge is defined by $Q = T_{3L}
+ \frac{1}{2} Y$. The model also includes two types of extra
fermions:

\begin{itemize}
\item Right handed neutrinos $\Psi_\nu^o = ( 3 , 1 , 1 , 0
)_R\;$ required to cancel anomalies\cite{T.Yanagida1979}, and

\item the $SU(2)_L$ singlet vector-like fermions:
\end{itemize}

 \begin{equation}
U_{L,R}^o= ( 1 , 3 , 1 , \frac{4}{3} )  \qquad , \qquad D_{L,R}^o
= ( 1 , 3 , 1 ,- \frac{2}{3} )  \label{vectorquarks} \end{equation}

\begin{equation}
N_{L,R}^o= ( 1 , 1 , 1 , 0 )
\qquad , \qquad E_{L,R}^o= ( 1 , 1 , 1 , -2 ) \, ,\label{vectorleptons}\end{equation}

\noindent which conserve the previous anomaly cancellation. The transformation of these
vector-like fermions allows the mass invariant mass terms

\begin{equation}
M_U \:\bar{U}_L^o \:U_R^o \,+\, M_D \:\bar{D}_L^o \:D_R^o \,+\, M_E \:\bar{E}_L^o \:E_R^o + h.c. \;,
\end{equation}

\noindent and

\begin{equation}
m_D \,\bar{N}_L^o \,N_R^o \,+\, m_L \,\bar{N}_L^o\, (N_L^o)^c \,+\, m_R \,\bar{N}_R^o\, (N_R^o)^c \,+\,  h.c
\end{equation}

\vspace{2mm}
\noindent These $SU(2)_L$ weak singlets
vector-like fermions have been introduced to give masses at tree
level only to the third family of known fermions through Dirac
See-saw mechanisms. $M_U\;,M_D\;,M_E$ play a crucial role
to implement a hierarchical spectrum for quarks and charged lepton
masses and mixing, meanwhile $m_D\;,m_L\;,m_R$ play a similar role for neutrino masses and
lepton mixing, all together with the radiative corrections.

\section{$SU(3)$ family symmetry breaking}

The corresponding $SU(3)$ gauge bosons are defined  through their couplings to
fermions as

\begin{multline}
i {\cal{L}}_{int} = \frac{g_{H}}{2}
\left( \bar{f_1^o} \gamma_{\mu} f_1^o- \bar{f_2^o} \gamma_{\mu} f_2^o \right) Z_1^\mu
+  \frac{g_{H}}{2 \sqrt{3}} \left( \bar{f_1^o} \gamma_{\mu} f_1^o+ \bar{f_2^o}
\gamma_{\mu} f_2^o - 2 \bar{f_3^o}
\gamma_{\mu} f_3^o  \right) Z_2^\mu                \\
+ \frac{g_{H}}{\sqrt{2}} \left( \bar{f_1^o} \gamma_{\mu} f_2^o \,Y_1^{+} +
\bar{f_1^o} \gamma_{\mu} f_3^o \,Y_2^{+} + \bar{f_2^o} \gamma_{\mu} f_3^o \,Y_3^{+} + h.c. \right) \label{SU3lagrangian}
\end{multline}

\noindent $f_1^o=u^o, d^o, e^o, \nu_e^o \; , \; f_2^o=c^o, s^o, \mu^o, \nu_\mu^o$  and $f_3^o=t^o, b^o, \tau^o, \nu_\tau^o$. To implement a hierarchical spectrum for charged fermion masses,
and simultaneously to achieve the SSB of $SU(3)$, we introduce the
flavon scalar fields: $\eta_i=(3 , 1 , 1 , 0),\;i=1,2,3$, transforming as the fundamental representation
under $SU(3)$ and being standard model singlets, with the "Vacuum Expectation
Values" (VEV's):

\begin{equation}
\langle \eta_1 \rangle^T = ( \Lambda_1 , 0 , 0) \quad , \quad
\langle \eta_2 \rangle^T = ( 0 , \Lambda_2 , 0) \quad , \quad
\langle \eta_3 \rangle^T = ( 0 , 0,  \Lambda_3)  \:. \label{veveta2eta3} \end{equation}

\noindent \emph{Actually, let us point out here that only two scalar flavons in the fundamental representation are
needed to completely break down the $SU(3)$ symmetry. The most convenient way to accomplish the spontaneous
breaking of the $SU(3)$ family symmetry is under study}. Thus, the contribution to the horizontal gauge boson masses from Eq.(\ref{veveta2eta3}) read

\begin{itemize}
\item $\eta_1:\quad \frac{g_{H_1}^2 \Lambda_1^2}{2} ( Y_1^+ Y_1^- + Y_2^+ Y_2^-) +  \frac{g_{H_1}^2 \Lambda_1^2}{4} ( Z_1^2 + \frac{Z_2^2}{3} + 2 Z_1 \frac{Z_2}{ \sqrt{3}} ) $

\item $\eta_2:\quad \frac{g_{H_2}^2 \Lambda_2^2}{2} ( Y_1^+ Y_1^- + Y_3^+ Y_3^-) +  \frac{g_{H_2}^2 \Lambda_2^2}{4} ( Z_1^2 + \frac{Z_2^2}{3} - 2 Z_1 \frac{Z_2}{ \sqrt{3}} ) $

\item $\eta_3:\quad \frac{g_{H_3}^2 \Lambda_3^2}{2} ( Y_2^+ Y_2^- + Y_3^+ Y_3^-) + g_{H_3}^2 \Lambda_3^2
\frac{Z_2^2}{3} $
\end{itemize}

\noindent \emph{Therefore, neglecting tiny contributions from electroweak symmetry breaking},
we obtain the gauge boson mass terms

\begin{multline}
( M_1^2 + M_2^2) \,Y_1^+ Y_1^- + ( M_1^2 + M_3^2) \,Y_2^+ Y_2^- + ( M_2^2 + M_3^2) \,Y_3^+ Y_3^- \\\\
+ \frac{1}{2}( M_1^2 + M_2^2 ) \,Z_1^2 +\frac{1}{2} \frac{M_1^2 + M_2^2+4 M_3^2}{3} \,Z_2^2
+ \frac{1}{2}( M_1^2 - M_2^2 ) \frac{2}{\sqrt{3}}  \,Z_1 \,Z_2
\end{multline}

\begin{equation} M_1^2=\frac{g_{H_1}^2 \Lambda_1^2}{2} \qquad , \qquad  M_2^2= \frac{g_{H_2}^2 \Lambda_2^2}{2} \qquad , \qquad M_3^2=\frac{g_{H_3}^2 \Lambda_3^2}{2}
\label{M1M2} \end{equation}

\begin{table}[!]
\begin{center} \begin{tabular}{ c | c c }
   &  $Z_1$ & $Z_2$ \\
\hline    \\
$Z_1$ &   $M_1^2 + M_2^2$ &  $ \frac{M_1^2 - M_2^2}{\sqrt{3}}$ \\
      &           &                            \\
$Z_2$  & $\frac{M_1^2 - M_2^2}{\sqrt{3}}$  & $\quad \frac{M_1^2 + M_2^2+4 M_3^2}{3}$
\end{tabular} \end{center}
\caption{$Z_1 - Z_2$ mixing mass matrix }
\end{table}

\noindent From the diagonalization of the $Z_1-Z_2$ squared mass matrix, we obtain
the eigenvalues

\begin{eqnarray*}
M_-^2=\frac{2}{3} \left( M_1^2 + M_2^2 + M_3^2 - \sqrt{ (M_2^2 -  M_1^2)^2 + (M_3^2 -  M_1^2)(M_3^2 -  M_2^2) } \right)                                        \\\\
M_+^2=\frac{2}{3} \left( M_1^2 + M_2^2 + M_3^2 + \sqrt{ (M_2^2 -  M_1^2)^2 + (M_3^2 -  M_1^2)(M_3^2 -  M_2^2) } \right)
\label{MmMp} \end{eqnarray*}

\begin{equation}
 M_{Y_1}^2\,Y_1^+ Y_1^- + M_{Y_2}^2\,Y_2^+ Y_2^- +  M_{Y_3}^2\,Y_3^+ Y_3^-
+ M_-^2 \,\frac{Z_-^2}{2} +  M_+^2 \,\frac{Z_+^2}{2}
\end{equation}

\noindent where

\begin{equation}
M_{Y_1}^2=M_1^2+M_2^2 \quad,\quad M_{Y_2}^2=M_1^2+M_3^2 \quad,\quad M_{Y_3}^2=M_2^2+M_3^2 \,
\end{equation}

\begin{equation}
\begin{pmatrix} Z_1 \\ Z_2  \end{pmatrix} = \begin{pmatrix} \cos\phi & - \sin\phi \\
\sin\phi & \cos\phi  \end{pmatrix} \begin{pmatrix} Z_- \\ Z_+  \end{pmatrix} \label{z1z2mixing}
\end{equation}

\begin{equation*} \cos\phi \, \sin\phi=\frac{\sqrt{3}}{4} \,
\frac{M_1^2 - M_2^2}{\sqrt{ (M_2^2 -  M_1^2)^2 + (M_3^2 -  M_1^2)(M_3^2 -  M_2^2) } } \; ,
\end{equation*}

\noindent with $Z_-$, $Z_+$ the mass eigenfields\footnote{Notice that in the limit $M_1^2 = M_2^2; \quad \sin\phi=0, \,\cos\phi=1 $}, and the hierarchy $M_1 , M_2, M_3 \gg M_W$. Due to the $Z_1 - Z_2$ mixing we diagonalize the propagators involving  $Z_1$ and $Z_2$  gauge bosons according to Eq.(\ref{z1z2mixing}):

\begin{equation*}
Z_1 = \cos\phi \;Z_- - \sin\phi \;Z_+  \quad , \quad Z_2 = \sin\phi \;Z_- + \cos\phi \;Z_+
\end{equation*}

\begin{eqnarray*}
\langle Z_1 Z_1 \rangle  &=&  \cos^2\phi\; \langle Z_- Z_- \rangle  +  \sin^2\phi\; \langle Z_+ Z_+ \rangle \\\\
\langle Z_2 Z_2 \rangle  &=&  \sin^2\phi\; \langle Z_- Z_- \rangle  +  \cos^2\phi\; \langle Z_+ Z_+ \rangle \\\\
\langle Z_1 Z_2 \rangle  &=&  \cos\phi \, \sin\phi \;( \langle Z_- Z_- \rangle - \langle Z_+ Z_+ \rangle )
\end{eqnarray*}

\section{Electroweak symmetry breaking}

Recently ATLAS\cite{ATLAS} and CMS\cite{CMS} at the Large Hadron Collider announced
the discovery of a Higgs-like particle, whose properties, couplings to fermions
and gauge bosons will determine whether it is the SM Higgs or a member of an extended
Higgs sector associated to a BSM theory.  The electroweak symmetry breaking in the
$SU(3)$ family symmetry model involves the introduction of two triplets of $SU(2)_L$
Higgs doublets.

\vspace{2mm}
To achieve the spontaneous breaking of the electroweak symmetry to
$U(1)_Q$,  we introduce the scalars: $\Phi^u = ( 3 , 1 , 2 , -1 )$
and $\Phi^d = ( 3 , 1 , 2 , +1 )$, with the VEV´s:

\vspace{2mm}
\begin{equation}
\langle \Phi^u \rangle = \begin{pmatrix}  \langle \Phi_1^u \rangle \\ \langle \Phi_2^u \rangle \\ \langle \Phi_3^u \rangle \end{pmatrix}  \quad , \quad
\langle \Phi^d \rangle= \begin{pmatrix} \langle \Phi_1^d \rangle \\ \langle \Phi_2^d \rangle
\\ \langle \Phi_3^d \rangle \end{pmatrix} \;,
\end{equation}

\vspace{2mm}
\begin{equation}
\langle \Phi_i^u \rangle = \frac{1}{\sqrt[]{2}}
\begin{pmatrix} v_{ui} \\ 0  \end{pmatrix}  \quad , \quad
\langle \Phi_i^d \rangle = \frac{1}{\sqrt[]{2}}
\begin{pmatrix} 0 \\ v_{di}  \end{pmatrix}  \:,\end{equation}

\vspace{2mm}
\noindent contribute to the W and Z boson masses:
\begin{equation*} \frac{g^2 }{4} \,(v_u^2+v_d^2)\,
W^{+} W^{-} + \frac{ (g^2 + {g^\prime}^2) }{8}  \,(v_u^2+v_d^2)\,Z_o^2 \end{equation*}

\vspace{2mm}
\noindent $v_u^2=v_{u1}^2+v_{u2}^2+v_{u3}^2\;$ , $\;v_d^2=v_{d1}^2+v_{d2}^2+v_{d3}^2 $.  Hence,
if we define $M_W=\frac{1}{2} g\, v$, we may write $ v=\sqrt{v_u^2+v_d^2 } \thickapprox 246$ GeV.

\vspace{5mm}
\section{Tree level neutrino masses}

Now we describe briefly the procedure to get the masses for
ordinary fermions. The analysis for quarks and charged leptons has
already discussed in \cite{albinosu3bled}. Here, we introduce the procedure
for neutrinos.

\subsection{Tree level Dirac neutrino masses}

With the fields of particles introduced in
the model, we may write the Dirac type gauge invariant Yukawa couplings

\begin{equation} h_D \,\bar{\Psi}_l^o \,\Phi^u \,N_R^o\;\;+\;\;
h_1 \,\bar{\Psi}_\nu^o \,\eta_1 \,N_L^o \;\;+\;\;
h_2 \,\bar{\Psi}_\nu^o \,\eta_2 \,N_L^o \;\;+\;\;h_3 \,\bar{\Psi}_\nu^o \,\eta_3 \,N_L^o +\;\ M_D \,\bar{N}_L^o \,N_R^o \;\;+ h.c
\label{nutlDirac} \end{equation}

\noindent $h_D$, $h_1$, $h_2$ and $h_3$ are Yukawa couplings, and $M_D$  a Dirac type,
invariant neutrino mass for the sterile neutrinos $N_{L,R}^o$. After electroweak
symmetry breaking, we obtain in the interaction basis ${\Psi_\nu^o}_{L,R}^T = ( \nu_e^o , \nu_\mu^o , \nu_\tau^o , N^o )_{L,R}$,  the mass terms

\begin{equation} h_D \left[  v_ 1 \,\bar{\nu}_{e L}^o + v_ 2 \,\bar{\nu}_{\mu L}^o + v_ 3  \,\bar{\nu}_{\tau L}^o \right] N_R^o + \left[ h_1 \Lambda_1 \,\bar{\nu}_{e R}^o +  h_2 \Lambda_2 \,\bar{\nu}_{\mu R}^o + h_3 \Lambda_3 \,\bar{\nu}_{\tau R}^o \right] N_L^o  + M_D \,\bar{N}_L^o \,N_R^o + h.c.
\end{equation}

\subsection{Tree level Majorana masses:}

 Since $N_{L,R}^o$, Eq.(\ref{vectorleptons}), are completely sterile neutrinos, we may also write the left and right handed Majorana type couplings

\begin{equation} h_L \,\bar{\Psi}_l^o \,\Phi^u (N_L^o)^c  \; + \; m_L \,\bar{N}_L^o\, (N_L^o)^c + h.c \end{equation}

\noindent and

\begin{equation} h_{1 R} \,\bar{\Psi}_\nu^o \,\eta_1 \,(N_R^o)^c \;+\; h_{2 R} \,\bar{\Psi}_\nu^o \,\eta_2 \,(N_R^o)^c \;+\;h_{3 R} \,\bar{\Psi}_\nu^o \,\eta_3 \,(N_R^o)^c  \;+\;m_R \,\bar{N}_R^o \,(N_R^o)^c + h.c \; ,\end{equation}

\noindent respectively. After spontaneous symmetry breaking, we also get the left handed and right handed Majorana mass terms

\begin{equation} h_L \,\left[  v_ 1 \,\bar{\nu}_{e L}^o + v_ 2 \,\bar{\nu}_{\mu L}^o + v_ 3 \, \bar{\nu}_{\tau L}^o         \right] \,(N_L^o)^c  \; + \; m_L \,\bar{N}_L^o \,(N_L^o)^c  + h.c. \, ,
\label{nutlML} \end{equation}

\begin{equation}
\left[h_{1 R} \,\Lambda_1 \,\bar{\nu}_{e R}^o +  h_{2 R} \,\Lambda_2 \,\bar{\nu}_{\mu R}^o + h_{3 R} \,\Lambda_3 \,\bar{\nu}_{\tau R}^o  \right] \,(N_R^o)^c  \,+\,m_R \,\bar{N}_R^o \,(N_R^o)^c + h.c. \, ,
\label{nutlMR} \end{equation}

\vspace{2mm}

\begin{table}[h]
\centering
\begin{tabular}{ c | c c c c c c c c}
&$(\nu^o_{e L})^c$ & $(\nu^o_{\mu L})^c$ & $(\nu^o_{\tau L})^c$ & $(N^o_L)^c$ & $\nu^o_{e R}$ & $\nu^o_{\mu R}$ & $\nu^o_{\tau R}$ & $N_R^o$ \\
\hline
$\overline{\nu^o_{e L}} $  & 0 & 0 & 0 & $h_L\, v_1$ & 0 &  0 & 0  & $h_D\, v_1$  \\\\
$\overline{\nu^o_{\mu L} }$  & 0 & 0 & 0 & $h_L\, v_2$ & 0 &  0 & 0  & $h_D\,  v_2$  \\\\
$\overline{\nu^o_{\tau L} }$  & 0 & 0 & 0 & $h_L\, v_3$ & 0 & 0 & 0  & $h_D\,  v_3$  \\\\
$\overline{N^o_L}$  & $h_L\, v_1$  & $h_L\, v_2$ & $h_L\, v_3$ & $m_L$ &  $h_1\,  \Lambda_1$  & $h_2\,  \Lambda_2$ & $h_3\, \Lambda_3$  & $m_D$  \\\\
$\overline{(\nu^o_{e R})^c}$  & 0 & 0 & 0  & $h_1\, \Lambda_1$ & 0 & 0 & 0 & $h_{1 R}\, \Lambda_1$     \\\\
$\overline{(\nu^o_{\mu R})^c}$  & 0 & 0 & 0  & $h_2\, \Lambda_2$ & 0 &  0 & 0 & $h_{2 R}\, \Lambda_2$  \\\\
$\overline{(\nu^o_{\tau R})^c}$  & 0 & 0 & 0 & $h_3\, \Lambda_3$ & 0 &  0 & 0 & $h_{3 R}\, \Lambda_3$  \\\\
$\overline{(N^o_R)^c}$ & $h_D\,  v_1$  & $h_D\,  v_2$  & $h_D\,  v_3$ & $m_D$  & $h_{1 R}\, \Lambda_1$  & $h_{2 R}\, \Lambda_2$  & $h_{3 R}\, \Lambda_3$  & $m_R$
\end{tabular}
\caption{Tree Level Majorana masses }
\label{treelevelMajorana} \end{table}

\vspace{2mm}
\noindent Thus, in the basis

\begin{equation}
{\Psi_\nu^o}^T= \left( \,\nu^o_{e L} \, , \, \nu^o_{\mu L} \, , \,  \nu^o_{\tau L} \, , \,  N^o_L \, , \,  (\nu^o_{e R})^c \, , \, (\nu^o_{\mu R})^c \, , \, (\nu^o_{\tau R})^c \, , \, (N^o_R)^c \, \right) \, ,\label{intbasis} \end{equation}

\noindent the Generic $8\times 8$ tree level Majorana mass matrix for neutrinos $\mathcal{M}_\nu^o$, from Table \ref{treelevelMajorana}, $\bar{\Psi_\nu^o}\; \mathcal{M}_\nu^o \;(\Psi_\nu^o)^c $, read

\begin{equation}
\mathcal{M}_\nu^o=
\begin{pmatrix}
\mathcal{M}_L^o & \mathcal{M}_D^o  \\\\ \mathcal{M}_D^{o\,T} & \mathcal{M}_R^o
\end{pmatrix} \label{nuoMajorana}
\end{equation}

\vspace{2mm} \noindent where

\begin{equation}
\mathcal{M}_L^o= \begin{pmatrix}
0 & 0 & 0 & \alpha_1 \\
0 & 0 & 0 & \alpha_2 \\
0 & 0 & 0 & \alpha_3 \\
\alpha_1  & \alpha_2 & \alpha_3 & m_L
\end{pmatrix} \quad , \quad
\mathcal{M}_R^o= \begin{pmatrix}
0 &  0 & 0 & \beta_1  \\
0 &  0 & 0  & \beta_2 \\
0 &  0 & 0  & \beta_3  \\
\beta_1 & \beta_2  & \beta_3  & m_R
\end{pmatrix} \quad , \quad
\mathcal{M}_D^o= \begin{pmatrix}
0 &  0 & 0  & a_1  \\
0 &  0 & 0  & a_2 \\
0 & 0 & 0  & a_3 \\
b_1 & b_2 & b_3  & m_D
\end{pmatrix}  \; ,
\end{equation}

\begin{equation} \alpha_i=h_L\, v_i \quad , \quad a_i=h_D\, v_i  \quad , \quad b_i=h_i\, \Lambda_i \quad , \quad
\beta_i=h_{i R}\,\Lambda_i
\end{equation}

\vspace{2mm}
\noindent Diagonalization of $\mathcal{M}_\nu^{(o)}$, Eq.(\ref{nuoMajorana}), yields four zero eigenvalues,
associated to the neutrino fields:

\begin{equation}
\frac{a_2}{ap}\,\nu^o_{e L}-\frac{a_1}{ap}\,\nu^o_{\mu L} \quad ,\quad
\frac{a_1 \,a_3}{ap \,a}\,\nu^o_{e L}+\frac{a_2 \,a_3}{ap \,a}\,\nu^o_{\mu L}-\frac{a_p}{a}\,\nu^o_{\tau L}, \label{masslessnuL}
\end{equation}

\begin{equation}
\frac{b_2}{bp}\,\nu^o_{e R}-\frac{b_1}{bp}\,\nu^o_{\mu R} \quad ,\quad
\frac{b_1 \,b_3}{bp \,b}\,\nu^o_{e R}+\frac{b_2 \,b_3}{bp \,b}\,\nu^o_{\mu R}-\frac{b_p}{b}\,\nu^o_{\tau R}\, ,\label{masslessnuR}
\end{equation}

\vspace{2mm}
\begin{equation*}
ap=\sqrt{a_1^2+a_2^2} \;,\; bp=\sqrt{b_1^2+b_2^2} \;,\; a=\sqrt{a_1^2+a_2^2+a_3^2}  \;,\;
b=\sqrt{b_1^2+b_2^2+b_3^2} \;.
\end{equation*}

\noindent Assuming for simplicity  $\frac{h_{1R}}{h_1}=\frac{h_{2R}}{h_2}=\frac{h_{3R}}{h_3}\equiv c_R$, that is

\begin{equation*}
\frac{\alpha_i}{a_i}=\frac{h_L}{h_D}=c_L \quad , \quad \frac{\beta_i}{b_i}=\frac{h_{iR}}{h_i}=c_R\, ,
\end{equation*}

\noindent the Characteristic Polynomial for the
nonzero eigenvalues of $\mathcal{M}_\nu^o$ reduce to the one of the matrix
$m_4$\footnote{The relation $a\,b=\alpha\,\beta$ would yield five massless neutrinos at tree level.}, Eq.(\ref{m4}),
where

\vspace{2mm}
\begin{equation}  m_4=
\begin{pmatrix}
0 & \alpha & 0  & a  \\
\alpha & m_L  & b   & m_D  \\
0 & b & 0  & \beta  \\
a & m_D & \beta   & m_R
\end{pmatrix} \quad , \quad  U_4=
\begin{pmatrix}
u_{11} & u_{12} & u_{13} & u_{14} \\
u_{21} & u_{22} & u_{23} & u_{24} \\
u_{31} & u_{32} & u_{33} & u_{34} \\
u_{41} & u_{42} & u_{43} & u_{44}
\end{pmatrix} \label{m4} \end{equation}

\begin{equation*}
 \alpha=\sqrt{\alpha_1^2+\alpha_2^2+\alpha_3^2}  \quad , \quad
\beta=\sqrt{\beta_1^2+\beta_2^2+\beta_3^2} \;.
\end{equation*}

\vspace{2mm}
\begin{equation} U_4^T\, m_4\, U_4 = Diag(m_5^o, m_6^o, m_7^o, m_8^o)\equiv d_4  \quad , \quad
m_4= U_4\,d_4\,U_4^T
\label{U4} \end{equation}

\vspace{2mm}
\noindent Eq.(\ref{U4}) impose the constrains

\begin{eqnarray}
u_{11}^2\,m_5^o+u_{12}^2\,m_6^o+u_{13}^2\,m_7^o+u_{14}^2\,m_8^o&=&0 \\ \nonumber\\
u_{31}^2\,m_5^o+u_{32}^2\,m_6^o+u_{33}^2\,m_7^o+u_{34}^2\,m_8^o&=&0 \\ \nonumber\\
u_{11} u_{31} \,m_5^o+u_{12} u_{32}\,m_6^o+u_{13} u_{33}\,m_7^o+u_{14} u_{34}\,m_8^o&=&0\,,
\end{eqnarray}

\vspace{2mm}
\noindent corresponding to the $(m_4)_{11}=(m_4)_{33}=(m_4)_{13}=0$ zero entries, respectively.

\vspace{4mm}
\noindent Therefore, $\mathcal{M}_\nu^o$ is  diagonalized by the orthogonal matrix

\vspace{4mm}
\begin{equation}  U_\nu^o=
\begin{pmatrix}
 \frac{a_2}{ap} & \frac{a_1 \,a_3}{a \,ap} & 0 & 0 & \frac{a_1}{a} u_{11} & \frac{a_1}{a} u_{12} & \frac{a_1}{a} u_{13} & \frac{a_1}{a} u_{14} \\\\
 - \frac{a_1}{ap} & \frac{a_2 \,a_3}{a \,ap} &  0 & 0 & \frac{a_2}{a} u_{11} & \frac{a_2}{a} u_{12} & \frac{a_2}{a} u_{13} & \frac{a_2}{a} u_{14} \\\\
0 & -\frac{ap}{a}  & 0 & 0 & \frac{a_3}{a} u_{11} & \frac{a_3}{a} u_{12} & \frac{a_3}{a} u_{13} & \frac{a_3}{a} u_{14} \\\\
 0 & 0 & 0 & 0 & u_{21} & u_{22} & u_{23} & u_{24} \\\\
0 & 0 & \frac{b_2}{bp} & \frac{b_1 \,b_3}{b \,bp} & \frac{b_1}{b} u_{31} & \frac{b_1}{b} u_{32} & \frac{b_1}{b} u_{33} & \frac{b_1}{b} u_{34}    \\\\
 0 & 0 & - \frac{b_1}{bp} & \frac{b_2 \,b_3}{b \,bp}  & \frac{b_2}{b} u_{31} & \frac{b_2}{b} u_{32} & \frac{b_2}{b} u_{33} & \frac{b_2}{b} u_{34} \\\\
0 & 0 & 0 & -\frac{b_p}{b} & \frac{b_3}{b} u_{31} & \frac{b_3}{b} u_{32} & \frac{b_3}{b} u_{33} & \frac{b_3}{b} u_{34} \\\\
0 & 0 & 0 & 0 & u_{41} & u_{42} & u_{43} & u_{44}
\end{pmatrix} \end{equation}

\vspace{5mm}
\begin{equation} (U_\nu^o)^T\,\mathcal{M}_\nu^o \,U_\nu^o   = Diag(0,0,0,0, m_5^o, m_6^o, m_7^o, m_8^o) \end{equation}

\section{One loop neutrino masses}

After tree level contributions light quarks, charged leptons\cite{albinosu32004,albinosu3bled} and
the two L-handed, Eq.(\ref{masslessnuL}) and  two R-handed, Eq.(\ref{masslessnuR}), neutrinos remain massless. So,
the initial fermion global symmetry, Eq.(\ref{globalsymmetry}), is broken down to
\begin{equation}
SU(2)_{q_L}\otimes SU(2)_{u_R}\otimes SU(2)_{d_R}\otimes
SU(2)_{l_L}\otimes SU(2)_{\nu_R}\otimes SU(2)_{e_R}\;.
\end{equation}

\noindent Therefore, in this scenario light neutrinos may get extremely small masses from
radiative corrections mediated by the $SU(3)$ heavy gauge bosons.

\vspace{5mm}
\subsection{One loop Dirac Neutrino masses}

Neutrinos may get tiny Dirac mass terms from the generic one loop diagram in Fig. 1, The internal fermion line
in this diagram represent the tree level see-saw mechanisms, Eqs.(\ref{nutlDirac}-\ref{nutlMR}). The vertices
read from the $SU(3)$ family symmetry interaction Lagrangian

\begin{multline}
i {\cal{L}}_{int} = \frac{g_{H}}{2}
\left( \bar{\nu_e^{o}} \gamma_{\mu} \nu_e^{o}- \bar{\nu_\mu^{o}} \gamma_{\mu} \nu_\mu^{o} \right) Z_1^\mu
+  \frac{g_{H}}{2 \sqrt{3}} \left( \bar{\nu_e^{o}} \gamma_{\mu} \nu_e^{o}+ \bar{\nu_\mu^{o}}
\gamma_{\mu} \nu_\mu^{o} - 2 \bar{\nu_\tau^{o}}
\gamma_{\mu} \nu_\tau^{o}  \right) Z_2^\mu                \\
+ \frac{g_{H}}{\sqrt{2}} \left( \bar{\nu_e^{o}} \gamma_{\mu} \nu_\mu^{o} \,Y_1^{+} +
\bar{\nu_e^{o}} \gamma_{\mu} \nu_\tau^{o} \,Y_2^{+} + \bar{\nu_\mu^{o}} \gamma_{\mu} \nu_\tau^{o} \,Y_3^{+} + h.c. \right) \label{SU3lagrangian}
\end{multline}

\noindent The contribution from these diagrams may be written as

\begin{equation} c_Y \frac{\alpha_H}{\pi}\,m_\nu(M_Y)_{ij} \quad ,\quad \alpha_H = \frac{g_H^2}{4 \pi} \, , \end{equation}

\noindent
\begin{equation}m_\nu(M_Y)_{ij} \equiv \sum_{k=5,6,7,8} m_k^o \:U^o_{ik} U^o_{jk}\:f(M_Y, m_k^o) \, , \end{equation}

\noindent $f(M_Y, m_k^o)=\frac{M_Y^2}{M_Y^2 - m_k^{o\,2}} \,ln{\frac{M_Y^2}{m_k^{o\,2}}} $ ,

\begin{figure}[!]
\centering\includegraphics[width=.9\textwidth]{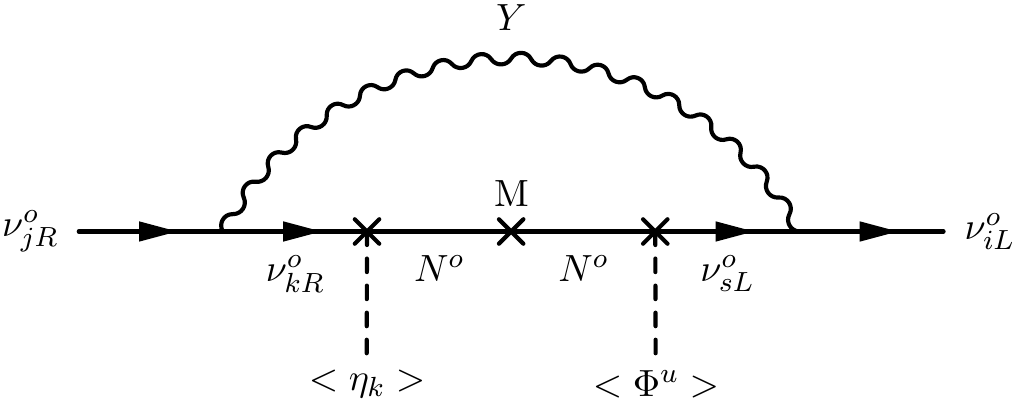}
\caption{ Generic one loop diagram contribution to the Dirac mass term
$m_{ij} \:{\bar{\nu}}_{iL}^o \nu_{jR}^o$. $\; \text{M}=M_D , m_L, m_R$}
\end{figure}

\vspace{5mm}
\begin{table}[h]
\begin{center}
\begin{tabular}{ c | c c c c }
   &  $\nu^o_{e R}$ & $\nu^o_{\mu R} $ & $\nu^o_{\tau R} $ & $N^o_R $ \\
\hline
$\bar{\nu}^o_{e L}$  & $D_{\nu \,11}$  & $D_{\nu \,12}$  & $D_{\nu \,13}$   & 0  \\
                                                                        \\
$\bar{\nu}^o_{\mu L}$  & $D_{\nu \,21}$ & $D_{\nu \,22}$  & $D_{\nu \,23}$    & 0  \\
                                                                          \\
$\bar{\nu}^o_{\tau L}$  & $D_{\nu \,31}$  &  $D_{\nu \,32}$  & $D_{\nu \,33}$ & 0   \\
                                                                           \\
$\bar{N}^o_L$  &  0 & 0 & 0 & 0
\end{tabular} \end{center}
\caption{One loop Dirac mass terms $ \frac{\alpha_H}{\pi}\,D_{\nu \,ij}   \,{\bar{\nu}}_{iL}^o \;\nu_{jR}^o$ }
\end{table}

\begin{equation} m_\nu(M_Y)_{i,4+j} = \frac{a_i\, b_j}{a\, b} \mathcal{F}_\nu (M_Y) \end{equation}

\begin{equation}
\mathcal{F}_\nu (M_Y)= \,u_{11} u_{31} \, m_5^o \,f(M_Y, m_5^o) + u_{12} u_{32} \,m_6^o \,f(M_Y, m_6^o) + u_{13} u_{33} \,m_7^o \,f(M_Y, m_7^o) + u_{14} u_{34} \,m_8^o \,f(M_Y, m_8^o)
\end{equation}

\begin{align*}
D_{\nu \,11} = & \;\frac{a_1 b_1}{a b} \left[ \frac{1}{4}\mathcal{F}_\nu (M_{Z_1})+\frac{1}{12}\mathcal{F}_\nu (M_{Z_2}) + \mathcal{F}_{\nu , m} \right]   + \frac{1}{2} \left[ \frac{a_2 b_2}{a b} \mathcal{F}_\nu (M_{Y_1}) + \frac{a_3 b_3}{a b} \mathcal{F}_\nu (M_{Y_2})       \right]  \; , \\\\
D_{\nu \,12} = & \; \frac{a_1 b_2}{a b} \,\left[ -\frac{1}{4} \mathcal{F}_\nu (M_{Z_1}) + \frac{1}{12}
\mathcal{F}_\nu (M_{Z_2})       \right]    \; , \\\\
D_{\nu \,13} = & \;\frac{a_1 b_3}{a b} \left[  - \frac{1}{6} \mathcal{F}_\nu (M_{Z_2})  - \mathcal{F}_{\nu , m}  \right]   \; , \\\\
D_{\nu \,21} = & \; \frac{a_2 b_1}{a b} \,\left[ -\frac{1}{4} \mathcal{F}_\nu (M_{Z_1}) + \frac{1}{12}
\mathcal{F}_\nu (M_{Z_2})       \right]    \; ,
\end{align*}
\begin{align*}
D_{\nu \,22} = & \;\frac{a_2 b_2}{a b} \,\left[ \frac{1}{4} \mathcal{F}_\nu (M_{Z_1}) + \frac{1}{12}
\mathcal{F}_\nu (M_{Z_2}) - \mathcal{F}_{\nu , m} \right]  +  \frac{1}{2} \left[ \frac{a_1 b_1}{a b}
\mathcal{F}_\nu (M_{Y_1}) + \frac{a_3 b_3}{a b} \mathcal{F}_\nu (M_{Y_3})   \right]     \; ,\\\\
D_{\nu \,23} = & \;\frac{a_2 b_3}{a b} \left[  - \frac{1}{6} \mathcal{F}_\nu (M_{Z_2})  + \mathcal{F}_{\nu , m}  \right]    \; ,\\\\
D_{\nu \,31} = & \;\frac{a_3 b_1}{a b} \left[  - \frac{1}{6} \mathcal{F}_\nu (M_{Z_2})  - \mathcal{F}_{\nu , m} \right]   \; , \\\\
D_{\nu \,32} = & \;\frac{a_3 b_2}{a b} \left[  - \frac{1}{6} \mathcal{F}_\nu (M_{Z_2})  + \mathcal{F}_{\nu , m} \right]    \; , \\\\
D_{\nu \,33} = & \;\frac{1}{3} \frac{a_3 b_3}{a b} \mathcal{F}_\nu (M_{Z_2}) + \frac{1}{2} \left[ \frac{a_1 b_1}{a b} \mathcal{F}_\nu (M_{Y_2}) + \frac{a_2 b_2}{a b} \mathcal{F}_\nu (M_{Y_3}) \right]      \; ,
\end{align*}

\begin{align*}
\mathcal{F}_\nu (M_{Z_1})=\, & \cos^2\phi \,\mathcal{F}_\nu (M_-) + \sin^2\phi \,\mathcal{F}_\nu (M_+)    \\\\
\mathcal{F}_\nu (M_{Z_2})=\, & \sin^2\phi \,\mathcal{F}_\nu (M_-) + \cos^2\phi \,\mathcal{F}_\nu (M_+)
\end{align*}

\begin{equation}
\mathcal{F}_{\nu , m} = \frac{1}{2\sqrt{3}}
\,\cos\phi \,\sin\phi \,[\mathcal{F}_\nu (M_- ) - \mathcal{F}_\nu (M_+ )]\;, \label{fmix}
\end{equation}

\vspace{5mm}
\subsection{One loop L-handed Majorana masses}

Neutrinos also obtain one loop corrections to L-handed and R-handed Majorana masses from
the diagrams of Fig. 2 and Fig. 3, respectively. A similar procedure as for Dirac Neutrino masses
leads to the one loop Majorana mass terms

\begin{equation} m_\nu(M_Y)_{i,j} = \frac{a_i\, a_j}{a^2} \mathcal{G}_\nu (M_Y) \end{equation}

\begin{equation}
\mathcal{G}_\nu(M_Y)=\,  m_5^o \,u_{11}^2 \,f(M_Y, m_5^o) + m_6^o \,u_{12}^2 \,f(M_Y, m_6^o) + m_7^o \,u_{13}^2 \,f(M_Y, m_7^o) + m_8^o \,u_{14}^2 \,f(M_Y, m_8^o)
\end{equation}

\begin{figure}[!]
\centering\includegraphics[width=.9\textwidth]{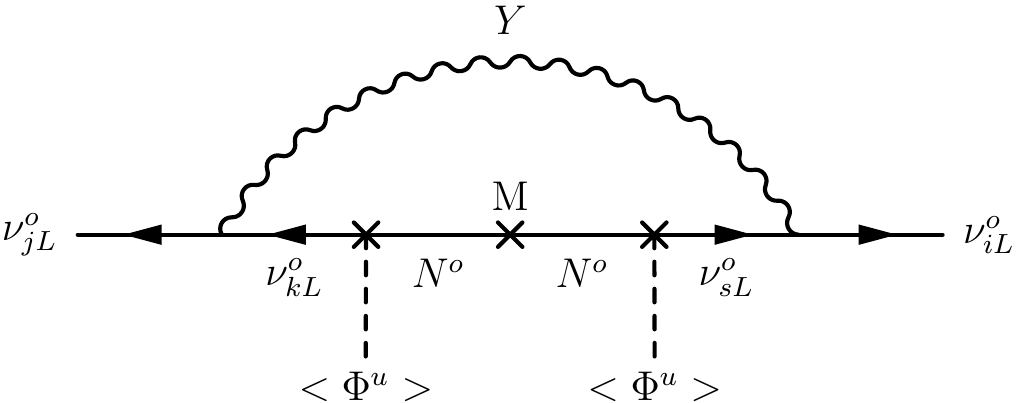}
\caption{ Generic one loop diagram contribution to the L-handed Majorana mass term
$m_{ij} \:{\bar{\nu}}_{iL}^o (\nu_{jL}^o)^T$. $\; \text{M}=M_D , m_L, m_R$}
\end{figure}

\vspace{5mm}
\begin{table}[h]
\begin{center}
\begin{tabular}{ c | c c c c }
   &  $\nu^o_{e L}$ & $\nu^o_{\mu L} $ & $\nu^o_{\tau L} $ & $N^o_L $ \\
\hline
$\nu^o_{e L}$  &  $L_{\nu \,11}$ & $L_{\nu \,12}$ &  $L_{\nu \,13}$   & 0  \\
                                                                        \\
$\nu^o_{\mu L}$  &  $L_{\nu \,12}$ & $L_{\nu \,22}$  & $L_{\nu \,23}$   & 0  \\
                                                                          \\
$\nu^o_{\tau L}$  & $L_{\nu \,13}$  & $L_{\nu \,23}$  & $L_{\nu \,33}$   & 0   \\
                                                                           \\
$N^o_L$  &  0 &\quad 0 &\quad 0 & 0
\end{tabular} \end{center}
\caption{One loop L-handed Majorana mass terms $\frac{\alpha_H}{\pi}\,L_{\nu \,ij}  \:{\bar{\nu}}_{iL}^o \;(\nu_{jL}^o)^T$ }
\end{table}

\begin{align*}
L_{\nu \,11} = & \; \frac{a_1^2}{a^2} \left[ \frac{1}{4} \mathcal{G}_\nu (M_{Z_1}) + \frac{1}{12} \mathcal{G}_\nu (M_{Z_2})       + \mathcal{G}_{\nu , m} \right]  \; , \\\\
L_{\nu \,22} = & \; \frac{a_2^2}{a^2} \left[ \frac{1}{4} \mathcal{G}_\nu (M_{Z_1}) + \frac{1}{12} \mathcal{G}_\nu (M_{Z_2})       - \mathcal{G}_{\nu , m} \right]  \; , \\\\
L_{\nu \,33} = & \;\frac{1}{3} \frac{a_3^2}{a^2}  \mathcal{G}_\nu (M_{Z_2})  \; , \\\\
L_{\nu \,12} = & \; \frac{a_1 a_2}{a^2} \left[ -\frac{1}{4} \mathcal{G}_\nu (M_{Z_1}) + \frac{1}{12} \mathcal{G}_\nu (M_{Z_2}) + \frac{1}{2} \mathcal{G}_\nu (M_1)   \right]  \; , \\\\
L_{\nu \,13} = & \; \frac{a_1 a_3}{a^2} \left[ -\frac{1}{6} \mathcal{G}_\nu (M_{Z_2}) + \frac{1}{2} \mathcal{G}_\nu (M_2)  - \mathcal{G}_{\nu , m} \right]  \; , \\\\
L_{\nu \,23} = & \; \frac{a_2 a_3}{a^2} \left[ -\frac{1}{6} \mathcal{G}_\nu (M_{Z_2}) + \frac{1}{2} \mathcal{G}_\nu (M_3)  + \mathcal{G}_{\nu , m} \right]
\end{align*}

\begin{align*}
\mathcal{G}_\nu (M_{Z_1})=\, & \cos^2\phi \,\mathcal{G}_\nu (M_-) + \sin^2\phi \,\mathcal{G}_\nu (M_+)    \\\\
\mathcal{G}_\nu (M_{Z_2})=\, & \sin^2\phi \,\mathcal{G}_\nu (M_-) + \cos^2\phi \,\mathcal{G}_\nu (M_+)
\end{align*}

\begin{equation}
\mathcal{G}_{\nu , m} =\,\frac{1}{2\sqrt{3}}
\,\cos\phi \,\sin\phi \,[\mathcal{G}_\nu (M_- ) - \mathcal{G}_\nu (M_+ )] \, , \label{gmix}
\end{equation}

\vspace{5mm}
\subsection{One loop R-handed Majorana masses}

\begin{equation} m_\nu(M_Y)_{4+i,4+j} = \frac{b_i\, b_j}{b^2} \mathcal{H}_\nu (M_Y) \end{equation}

\begin{equation}
\mathcal{H}_\nu(M_Y)=\,  m_5^o \,u_{31}^2 \,f(M_Y, m_5^o) + m_6^o \,u_{32}^2 \,f(M_Y, m_6^o) + m_7^o \,u_{33}^2 \,f(M_Y, m_7^o) + m_8^o \,u_{34}^2 \,f(M_Y, m_8^o)
\end{equation}

\begin{figure}[!]
\centering\includegraphics[width=.9\textwidth]{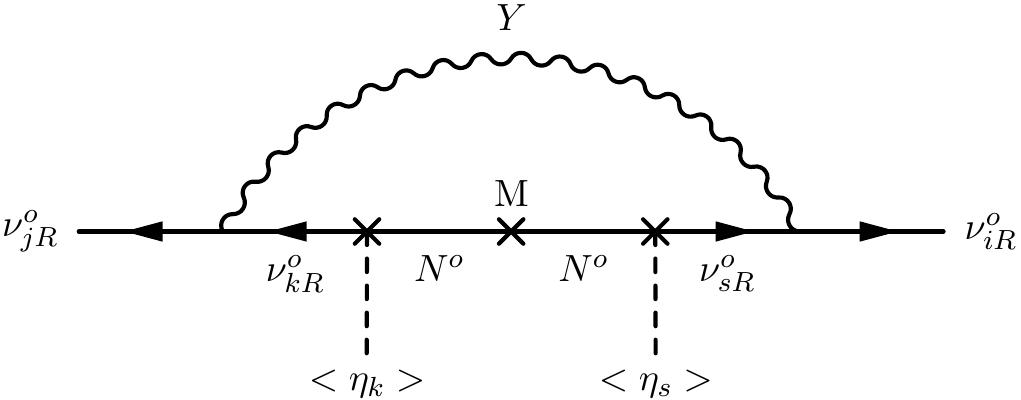}
\caption{ Generic one loop diagram contribution to the R-handed Majorana mass term
$m_{ij} \:{\bar{\nu}}_{iR}^o (\nu_{jR}^o)^T$. $\; \text{M}=M_D , m_L, m_R$}
\end{figure}

\vspace{5mm}
\begin{table}[h]
\begin{center}
\begin{tabular}{ c | c c c c }
   &  $\nu^o_{e R}$ & $\nu^o_{\mu R} $ & $\nu^o_{\tau R} $ & $N^o_R $ \\
\hline
$\nu^o_{e R}$  & $R_{\nu \,11}$   & $R_{\nu \,12}$  & $R_{\nu \,13}$  & 0  \\
                                                                        \\
$\nu^o_{\mu R}$  &  $R_{\nu \,12}$   & $R_{\nu \,22}$  & $R_{\nu \,23}$  & 0  \\
                                                                          \\
$\nu^o_{\tau R}$  & $R_{\nu \,13}$  & $R_{\nu \,23}$  & $R_{\nu \,33}$   & 0   \\
                                                                 \\
$N^o_R$  &  0 &\quad 0 &\quad 0 & 0
\end{tabular} \end{center}
\caption{One loop R-handed Majorana mass terms $ \frac{\alpha_H}{\pi}\,R_{\nu \,ij}  \:{\bar{\nu}}_{iR}^o \;(\nu_{jR}^o)^T$ }
\end{table}

\begin{align*}
R_{\nu \,11} = & \; \frac{b_1^2}{b^2} \left[ \frac{1}{4} \mathcal{H}_\nu (M_{Z_1}) + \frac{1}{12} \mathcal{H}_\nu (M_{Z_2})       + \mathcal{H}_{\nu , m} \right]  \; , \\\\
R_{\nu \,22} = & \; \frac{b_2^2}{b^2} \left[ \frac{1}{4} \mathcal{H}_\nu (M_{Z_1}) + \frac{1}{12} \mathcal{H}_\nu (M_{Z_2})       - \mathcal{H}_{\nu , m} \right]  \; , \\\\
R_{\nu \,33} = & \;\frac{1}{3} \frac{b_3^2}{b^2}  \mathcal{H}_\nu (M_{Z_2})  \; ,
\end{align*}

\begin{align*}
R_{\nu \,12} = & \; \frac{b_1 b_2}{b^2} \left[ -\frac{1}{4} \mathcal{H}_\nu (M_{Z_1}) + \frac{1}{12} \mathcal{H}_\nu (M_{Z_2}) + \frac{1}{2} \mathcal{H}_\nu (M_1)   \right]  \; , \\\\
R_{\nu \,13} = & \; \frac{b_1 b_3}{b^2} \left[ -\frac{1}{6} \mathcal{H}_\nu (M_{Z_2}) + \frac{1}{2} \mathcal{H}_\nu (M_2)  - \mathcal{H}_{\nu , m} \right]  \; , \\\\
R_{\nu \,23} = & \; \frac{b_2 b_3}{b^2} \left[ -\frac{1}{6} \mathcal{H}_\nu (M_{Z_2}) + \frac{1}{2} \mathcal{H}_\nu (M_3)  + \mathcal{H}_{\nu , m} \right]
\end{align*}

\begin{align*}
\mathcal{H}_\nu (M_{Z_1})=\, & \cos^2\phi \,\mathcal{H}_\nu (M_-) + \sin^2\phi \,\mathcal{H}_\nu (M_+)    \\\\
\mathcal{H}_\nu (M_{Z_2})=\, & \sin^2\phi \,\mathcal{H}_\nu (M_-) + \cos^2\phi \,\mathcal{H}_\nu (M_+)
\end{align*}

\begin{equation}
\mathcal{H}_{\nu , m} = \,\frac{1}{2\sqrt{3}}
\,\cos\phi \,\sin\phi \,[\mathcal{H}_\nu (M_- ) - \mathcal{H}_\nu (M_+ )] \;, \label{hmix}
\end{equation}

\vspace{4mm}

\noindent \emph{where $\mathcal{F}_{\nu , m}\;,\;\mathcal{G}_{\nu , m}$ and $\mathcal{H}_{\nu , m}$, Eqs.(\ref{fmix},\ref{gmix},\ref{hmix}), come from $Z_1-Z_2$ mixing diagram contributions.}

\vspace{4mm}
\noindent Thus, in the $\Psi_\nu^o$ basis, Eq.(\ref{intbasis}),  we may write the one loop contribution
for neutrinos $\bar{\Psi_\nu^o}\; \mathcal{M}_{1\,\nu}^o \;(\Psi_\nu^o)^c $,

\vspace{3mm}
\begin{equation} \mathcal{M}_{1\,\nu}^o=
\begin{pmatrix}
L_{\nu \,11} &\quad L_{\nu \,12} &\quad L_{\nu \,13} & \quad 0 &\quad D_{\nu \,11}  &\quad D_{\nu \,12}  &\quad D_{\nu \,13}   &\quad 0  \\\\
L_{\nu \,12} &\quad L_{\nu \,22}  &\quad L_{\nu \,23} & \quad 0&\quad D_{\nu \,21} &\quad D_{\nu \,22}  &\quad D_{\nu \,23}  &\quad 0   \\\\
L_{\nu \,13}  &\quad L_{\nu \,23} &\quad L_{\nu \,33} & \quad 0  &\quad D_{\nu \,31} &\quad D_{\nu \,32}  &\quad D_{\nu \,33}   &\quad 0 \\\\
0 &\quad 0 &\quad 0 &\quad 0 &\quad 0 &\quad 0 &\quad 0  &\quad 0  \\\\
D_{\nu \,11}  &\quad D_{\nu \,21} &\quad D_{\nu \,31} & \quad 0  &\quad R_{\nu \,11} &\quad R_{\nu \,12} &\quad R_{\nu \,13} &\quad 0  \\\\
D_{\nu \,12} &\quad D_{\nu \,22} &\quad D_{\nu \,32} &\quad 0 &\quad R_{\nu \,12} &\quad R_{\nu \,22} &\quad R_{\nu \,23} &\quad 0  \\\\
D_{\nu \,13} &\quad D_{\nu \,23} &\quad D_{\nu \,33} &\quad 0 &\quad R_{\nu \,13} &\quad R_{\nu \,23} &\quad R_{\nu \,33} &\quad 0  \\\\
0 &\quad 0 &\quad 0 &\quad 0 &\quad 0 &\quad 0 &\quad 0 &\quad 0
\end{pmatrix}   \, \frac{\alpha_H}{\pi} \end{equation}

\vspace{5mm}
\subsection{Neutrino mass matrix up to one loop}

Finally, we obtain the general symmetric Majorana mass matrix for neutrinos up to one loop

\begin{equation} \mathcal{M}_{\nu}= ( U_\nu^o )^T\, \mathcal{M}_{1\,\nu}^o \,U_\nu^o + Diag(0, 0, 0, 0, m_5^o, m_6^o, m_7^o, m_8^o) \,, \end{equation}

\noindent where explicitly

\begin{equation}
( U_\nu^o )^T\, \mathcal{M}_{1\,\nu}^o \,U_\nu^o =
\begin{pmatrix}
N_{11} \;&\; N_{12}  \;&\; N_{13}  \;&\; N_{14}  \;&\; N_{15}  \;&\; N_{16}  \;&\; N_{17}  \;&\; N_{18} \\\\
N_{12} \;&\; N_{22}  \;&\; N_{23}  \;&\; N_{24}  \;&\; N_{25}  \;&\; N_{26}  \;&\; N_{27}  \;&\; N_{28} \\\\
N_{13} \;&\; N_{23}  \;&\; N_{33}  \;&\; N_{34}  \;&\; N_{35}  \;&\; N_{36}  \;&\; N_{37}  \;&\; N_{38} \\\\
N_{14} \;&\; N_{24}  \;&\; N_{34}  \;&\; N_{44}  \;&\; N_{45}  \;&\; N_{46}  \;&\; N_{47}  \;&\; N_{48} \\\\
N_{15} \;&\; N_{25}  \;&\; N_{35}  \;&\; N_{45}  \;&\; N_{55}  \;&\; N_{56}  \;&\; N_{57}  \;&\; N_{58} \\\\
N_{16} \;&\; N_{26}  \;&\; N_{36}  \;&\; N_{46}  \;&\; N_{56}  \;&\; N_{66}  \;&\; N_{67}  \;&\; N_{68} \\\\
N_{17} \;&\; N_{27}  \;&\; N_{37}  \;&\; N_{47}  \;&\; N_{57}  \;&\; N_{67}  \;&\; N_{77}  \;&\; N_{78} \\\\
N_{18} \;&\; N_{28}  \;&\; N_{38}  \;&\; N_{48}  \;&\; N_{58}  \;&\; N_{68}  \;&\; N_{78}  \;&\; N_{88}
\end{pmatrix} \, \frac{\alpha_H}{\pi} \end{equation}

\vspace{3mm}
\noindent {\bf Majorana L-handed:}

\begin{align} N_{11}= & \frac{a_1^2 a_2^2}{a_p^2 a^2} (\mathcal{G}_{Z_1} - \mathcal{G}_1 ) \\ \nonumber\\
N_{12}= &- \frac{a_1 a_2 a_3}{2 a^3} [ \frac{a_2^2 - a_1^2}{a_p^2} (\mathcal{G}_{Z_1} - \mathcal{G}_1) + \mathcal{G}_2 - \mathcal{G}_3  - 6 \mathcal{G}_m  ]                                                      \\ \nonumber\\
N_{22}= &\frac{a_3^2}{a^2} \,\left[\, \frac{1}{4}\frac{(a_2^2 - a_1^2)^2}{a_p^2 \,a^2} (\mathcal{G}_{Z_1} - \mathcal{G}_1) + \frac{a_2^2}{a^2}(\mathcal{G}_2 - \mathcal{G}_3)   + \frac{a_p^2}{4 \,a^2}(\mathcal{G}_1+3 \mathcal{G}_{Z_2} - 4 \mathcal{G}_2) - 3 \frac{a_2^2 - a_1^2}{a^2} \mathcal{G}_m \,\right]
\end{align}

\vspace{5mm}
\noindent {\bf Dirac:}

\begin{align}
N_{13}= &\frac{1}{2 ap\, bp\, a\, b}\,
\left\{ (a_1^2 b_1^2+a_2^2 b_2^2)\mathcal{F}_1+a_3 b_3(a_2 b_2 \mathcal{F}_2+a_1 b_1 \mathcal{F}_3) + 2 a_1 b_1 a_2 b_2 \mathcal{F}_{Z_1} \right\}   \\ \nonumber\\
N_{14}= &\frac{1}{2 ap\, bp\, a\, b}  \frac{b_3}{b}\,
\left\{ b_1 b_2(a_2^2-a_1^2) \mathcal{F}_1+a_3 b_3(a_2 b_1 \mathcal{F}_2-a_1 b_2 \mathcal{F}_3) + a_1 a_2 (b_1^2-b_2^2)\mathcal{F}_{Z_1} + 6 a_1 a_2\, bp^2\, \mathcal{F}_m  \right\} \\ \nonumber\\
N_{23}= &\frac{1}{2 ap\, bp\, a\, b}  \frac{a_3}{a}\,
\left\{ a_1 a_2(b_2^2-b_1^2) \mathcal{F}_1+a_3 b_3(a_1 b_2 \mathcal{F}_2-a_2 b_1 \mathcal{F}_3) + b_1 b_2 (a_1^2-a_2^2)\mathcal{F}_{Z_1} + 6 b_1 b_2 \,ap^2\, \mathcal{F}_m   \right\} \\ \nonumber\\
N_{24}= &\frac{1}{ap\, bp\, a^2\, b^2}\,
\left\{ a_3 b_3[ a_1 b_1 a_2 b_2 \mathcal{F}_1 + \frac{1}{4}(a_1^2-a_2^2)(b_1^2-b_2^2) \mathcal{F}_{Z_1}+ \frac{3}{4} ap^2\, bp^2\, \mathcal{F}_{Z_2}] \right. \nonumber\\ & \left. \hspace{4cm} + \frac{1}{2} (a_3^2 b_3^2 + ap^2\, bp^2)(a_1 b_1 \mathcal{F}_2+a_2 b_2 \mathcal{F}_3) + 3 a_3 b_3( a_1^2 b_1^2 - a_2^2 b_2^2 ) \mathcal{F}_m \right\}
\end{align}

\pagebreak

\noindent {\bf Majorana R-handed:}

\begin{align} N_{33}= &\frac{b_1^2 b_2^2}{b_p^2 b^2} (\mathcal{H}_{Z_1} - \mathcal{H}_1 ) \\ \nonumber\\
N_{34}= &- \frac{b_1 b_2 b_3}{2 b^3} [ \frac{b_2^2 - b_1^2}{b_p^2} (\mathcal{H}_{Z_1} - \mathcal{H}_1) + \mathcal{H}_2 - \mathcal{H}_3  - 6 \mathcal{H}_m  ] \\ \nonumber\\
N_{44}= &\frac{b_3^2}{b^2} \,\left[\, \frac{1}{4}\frac{(b_2^2 - b_1^2)^2}{b_p^2 \,b^2} (\mathcal{H}_{Z_1} - \mathcal{H}_1) + \frac{b_2^2}{b^2}(\mathcal{H}_2 - \mathcal{H}_3) + \frac{b_p^2}{4 \,b^2}(\mathcal{H}_1+3 \mathcal{H}_{Z_2}  - 4 \mathcal{H}_2) - 3 \frac{b_2^2 - b_1^2}{b^2} \mathcal{H}_m \,\right]
\end{align}

\vspace{5mm}
\noindent {\bf Majorana L-handed and Dirac:}

\begin{eqnarray}
N_{15}=\mathcal{G}_{15}\,u_{11} + m_{13}\,u_{31} \qquad &; \qquad N_{16}=\mathcal{G}_{15}\,u_{12} + m_{13}\,u_{32}  \\
\nonumber \\
N_{17}=\mathcal{G}_{15}\,u_{13} + m_{13}\,u_{33} \qquad &; \qquad N_{18}=\mathcal{G}_{15}\,u_{14} + m_{13}\,u_{34}
\end{eqnarray}

\begin{equation*}
\mathcal{G}_{15}=-\frac{a_1\,a_2}{2\,a_p\,a} \,\left[\,\frac{a_2^2 - a_1^2}{a^2} (\mathcal{G}_{Z_1} - \mathcal{G}_1) + \frac{a_3^2}{a^2}(\mathcal{G}_3 - \mathcal{G}_2) + 2\, \frac{(2\,a_3^2 - a_p^2)}{a^2} \,\mathcal{G}_m \,\right]
\end{equation*}

\begin{equation*}
m_{13}= \frac{1}{2 ap\, a\, b^2}  \,
\left\{ b_1 b_2(a_2^2-a_1^2) \mathcal{F}_1+a_3 b_3(a_2 b_1 \mathcal{F}_2-a_1 b_2 \mathcal{F}_3)  + a_1 a_2 (b_1^2-b_2^2)\mathcal{F}_{Z_1}
+ 2 a_1 a_2 \,( bp^2 - 2 b_3^2) \mathcal{F}_m \right\}
\end{equation*}

\begin{eqnarray}
N_{25}=\mathcal{G}_{25}\,u_{11} + m_{23}\,u_{31} \qquad &; \qquad N_{26}=\mathcal{G}_{25}\,u_{12} + m_{23}\,u_{32}  \\
\nonumber \\
N_{27}=\mathcal{G}_{25}\,u_{13} + m_{23}\,u_{33} \qquad &; \qquad N_{28}=\mathcal{G}_{25}\,u_{14} + m_{23}\,u_{34}
\end{eqnarray}

\begin{multline*}
\mathcal{G}_{25}=\frac{a_3}{4\,a_p\,a^4}
\left\{ \,(a_2^2 - a_1^2)^2\,( \mathcal{G}_{Z_1} - \mathcal{G}_1) + 2\,a_2^2(a_3^2-a_p^2)\,(\mathcal{G}_3 - \mathcal{G}_2) - a_p^4\,(\mathcal{G}_{Z_2} - \mathcal{G}_1) \right. \\ \left.  - 2\,a_p^2(a_3^2-a_p^2)\,(\mathcal{G}_{Z_2} - \mathcal{G}_2) +
4\,(a_2^2 - a_1^2)\,(a_3^2 - 2 a_p^2) \,\mathcal{G}_m \, \right\}
\end{multline*}

\begin{multline*}
m_{23}= \frac{1}{ap\, a^2\, b^2}
\left\{ a_3[ a_1 b_1 a_2 b_2 \mathcal{F}_1 + \frac{1}{4}(a_1^2-a_2^2)(b_1^2-b_2^2) \mathcal{F}_{Z_1}+ \frac{1}{4} ap^2 (bp^2 -2 b_3^2) \mathcal{F}_{Z_2}] \right. \\ \left. + \frac{1}{2} b_3 (a_3^2 - ap^2)(a_1 b_1 \mathcal{F}_2+a_2 b_2 \mathcal{F}_3)
+ a_3 [ a_1^2(3 b_1^2-b^2) + a_2^2(b^2-3 b_2^2)] \mathcal{F}_m
\right\}
\end{multline*}

\vspace{5mm}
\noindent {\bf Dirac and Majorana R-handed:}

\begin{eqnarray}
N_{35}=m_{31}\,u_{11}+\mathcal{H}_{35}\,u_{31} \quad &, \quad N_{36}=m_{31}\,u_{12}+\mathcal{H}_{35}\,u_{32}  \\
                                                                                                    \nonumber\\
N_{37}=m_{31}\,u_{13}+\mathcal{H}_{35}\,u_{33} \quad &, \quad N_{38}=m_{31}\,u_{14}+\mathcal{H}_{35}\,u_{34}
\end{eqnarray}

\begin{equation*}
m_{31}= \frac{1}{2 bp\, a^2\, b}\,
\left\{ a_1 a_2(b_2^2-b_1^2) \mathcal{F}_1+a_3 b_3(a_1 b_2 \mathcal{F}_2-a_2 b_1 \mathcal{F}_3) + b_1 b_2 (a_1^2-a_2^2)\mathcal{F}_{Z_1}
+ 2 b_1 b_2 ( ap^2 - 2 a_3^2) \mathcal{F}_m  \right\} \end{equation*}

\begin{equation*}
\mathcal{H}_{35}=-\frac{b_1\,b_2}{2\,b_p\,b} \,\left[\,\frac{b_2^2 - b_1^2}{b^2} (\mathcal{H}_{Z_1} - \mathcal{H}_1) + \frac{b_3^2}{b^2}(\mathcal{H}_3 - \mathcal{H}_2) + 2\, \frac{(2\,b_3^2 - b_p^2)}{b^2} \,\mathcal{H}_m \,\right]
\end{equation*}

\vspace{3mm}

\begin{eqnarray}
N_{45}=m_{32}\,u_{11} + \mathcal{H}_{45}\,u_{31} \quad &, \quad N_{46}=m_{32}\,u_{12} + \mathcal{H}_{45}\,u_{32}  \\
\nonumber \\
N_{47}=m_{32}\,u_{13} + \mathcal{H}_{45}\,u_{33} \quad &, \quad N_{48}=m_{32}\,u_{14} + \mathcal{H}_{45}\,u_{34}
\end{eqnarray}

\begin{multline*}
m_{32}= \frac{1}{bp\,  a^2\, b^2}
\left\{ b_3[ a_1 b_1 a_2 b_2 \mathcal{F}_1 + \frac{1}{4}(a_1^2-a_2^2)(b_1^2-b_2^2) \mathcal{F}_{Z_1}+ \frac{1}{4} bp^2 ( ap^2-2 a_3^2 ) \mathcal{F}_{Z_2}] \right. \\ \left. + \frac{1}{2} a_3 (b_3^2 - bp^2)(a_1 b_1 \mathcal{F}_2+a_2 b_2 \mathcal{F}_3)
+ b_3 [ b_1^2(3 a_1^2-a^2) + b_2^2(a^2-3 a_2^2)] \mathcal{F}_m    \right\}
\end{multline*}

\begin{multline*}
\mathcal{H}_{45}=\frac{b_3}{4\,b_p\,b^4}
\left\{ \,(b_2^2 - b_1^2)^2\,( \mathcal{H}_{Z_1} - \mathcal{H}_1) + 2\,b_2^2(b_3^2-b_p^2)\,(\mathcal{H}_3 - \mathcal{H}_2) - b_p^4\,(\mathcal{H}_{Z_2} - \mathcal{H}_1) \right. \\ \left.  - 2\,b_p^2(b_3^2-b_p^2)\,(\mathcal{H}_{Z_2} - \mathcal{H}_2) +
4\,(b_2^2 - b_1^2)\,(b_3^2 - 2 b_p^2) \,\mathcal{H}_m \, \right\}
\end{multline*}

\vspace{5mm}
\noindent {\bf Majorana L-handed, Dirac and Majorana R-handed:}

\begin{eqnarray}
N_{55}&=&\mathcal{G}_{55}\,u_{11}^2 + 2\,m_{33}\,u_{11}\,u_{31} + \mathcal{H}_{55}\,u_{31}^2  \\ \nonumber\\
N_{56}&=&\mathcal{G}_{55}\,u_{11}\,u_{12} + m_{33}\,(u_{11}\,u_{32}+u_{12}\,u_{31}) + \mathcal{H}_{55}\,u_{31}\,u_{32}  \\ \nonumber\\
N_{57}&=&\mathcal{G}_{55}\,u_{11}\,u_{13} + m_{33}\,(u_{11}\,u_{33}+u_{13}\,u_{31}) + \mathcal{H}_{55}\,u_{31}\,u_{33}  \\ \nonumber\\
N_{58}&=&\mathcal{G}_{55}\,u_{11}\,u_{14} + m_{33}\,(u_{11}\,u_{34}+u_{14}\,u_{31}) + \mathcal{H}_{55}\,u_{31}\,u_{34}  \\ \nonumber\\
N_{66}&=&\mathcal{G}_{55}\,u_{12}^2 + 2\,m_{33}\,u_{12}\,u_{32} + \mathcal{H}_{55}\,u_{32}^2  \\ \nonumber\\
N_{67}&=&\mathcal{G}_{55}\,u_{12}\,u_{13} + m_{33}\,(u_{13}\,u_{32}+u_{12}\,u_{33}) + \mathcal{H}_{55}\,u_{32}\,u_{33}  \\ \nonumber\\
N_{68}&=&\mathcal{G}_{55}\,u_{12}\,u_{14} + m_{33}\,(u_{14}\,u_{32}+u_{12}\,u_{34}) + \mathcal{H}_{55}\,u_{32}\,u_{34}  \\ \nonumber\\
N_{77}&=&\mathcal{G}_{55}\,u_{13}^2 + 2\,m_{33}\,u_{13}\,u_{33} + \mathcal{H}_{55}\,u_{33}^2  \\ \nonumber\\
N_{78}&=&\mathcal{G}_{55}\,u_{13}\,u_{14} + m_{33}\,(u_{14}\,u_{33}+u_{13}\,u_{34}) + \mathcal{H}_{55}\,u_{33}\,u_{34}  \\ \nonumber\\
N_{88}&=&\mathcal{G}_{55}\,u_{14}^2 + 2\,m_{33}\,u_{14}\,u_{34} + \mathcal{H}_{55}\,u_{34}^2                                                \end{eqnarray}

\begin{equation*}
\mathcal{G}_{55}= \frac{a_1^2\,a_2^2}{a^4} \,\mathcal{G}_1 + \frac{a_1^2\,a_3^2}{a^4} \,\mathcal{G}_2 + \frac{a_2^2\,a_3^2}{a^4} \,\mathcal{G}_3 +
\frac{(a_2^2 - a_1^2)^2}{4\,a^4} \,\mathcal{G}_{Z_1}
+ \frac{(2 a_3^2 - a_p^2)^2}{12\,a^4} \,\mathcal{G}_{Z_2}
+ \frac{(a_2^2 - a_1^2)\,(2\,a_3^2 - a_p^2)}{a^4}\, \mathcal{G}_m
\end{equation*}

\begin{multline*}
m_{33}= \frac{1}{a^2\, b^2}
\left\{  a_1 b_1 a_2 b_2 \mathcal{F}_1 + \frac{1}{4}(a_1^2-a_2^2)(b_1^2-b_2^2) \mathcal{F}_{Z_1} + \frac{1}{12}( ap^2 - 2 a_3^2)(bp^2 - 2 b_3^2)  \mathcal{F}_{Z_2} \right. \\ \left. + a_3 b_3(a_1 b_1 \mathcal{F}_2+a_2 b_2 \mathcal{F}_3)
+ [ a_1^2 b_1^2 - a_2^2 b_2^2 + a_3^2( b_2^2 - b_1^2) + b_3^2( a_2^2 - a_1^2 ) ] \mathcal{F}_m   \right\}
\end{multline*}

\begin{equation*}
\mathcal{H}_{55}= \frac{b_1^2\,b_2^2}{b^4} \,\mathcal{H}_1 + \frac{b_1^2\,b_3^2}{b^4} \,\mathcal{H}_2 + \frac{b_2^2\,b_3^2}{b^4} \,\mathcal{H}_3 +
\frac{(b_2^2 - b_1^2)^2}{4\,b^4} \,\mathcal{H}_{Z_1}
+ \frac{(2 b_3^2 - b_p^2)^2}{12\,b^4} \,\mathcal{H}_{Z_2}
+ \frac{(b_2^2 - b_1^2)\,(2\,b_3^2 - b_p^2)}{b^4}\, \mathcal{H}_m
\end{equation*}

\vspace{5mm}
\subsection{Quark $( V_{CKM} )_{4\times 4}$ and Lepton $( U_{PMNS} )_{4\times 8}$  mixing matrices }

Within this $SU(3)$ family symmetry model, the transformation from
massless to physical mass fermion eigenfields for quarks and charged leptons is

\begin{equation*} \psi_L^o = V_L^{o} \:V^{(1)}_L \:\Psi_L \qquad \mbox{and}
\qquad \psi_R^o = V_R^{o} \:V^{(1)}_R \:\Psi_R \,,\end{equation*}

\noindent and for neutrinos $\Psi_\nu^o =  U_\nu^o \, U_\nu \,\Psi_\nu$. Recall now
that vector like quarks, Eq.(\ref{vectorquarks}), are $SU(2)_L$
weak singlets, and hence, they do not couple to $W$ boson in the
interaction basis. In this way, the interaction of  L-handed up and down quarks; ${f_{uL}^o}^T=(u^o,c^o,t^o)_L$ and
${f_{dL}^o}^T=(d^o,s^o,b^o)_L$, to the $W$ charged gauge boson is

\begin{equation} \frac{g}{\sqrt{2}} \,\bar{f^o}_{u L} \gamma_\mu f_{d L}^o
{W^+}^\mu = \frac{g}{\sqrt{2}} \,\bar{\Psi}_{u L}\;
[(V_{u L}^o\,V_{u L}^{(1)})_{3\times 4}]^T \;(V_{d L}^o\,V_{d L}^{(1)})_{3\times 4}\;
\gamma_\mu \Psi_{d L} \;{W^+}^\mu \:,\end{equation}

\noindent $g$ is the $SU(2)_L$ gauge coupling. Hence, the non-unitary $V_{CKM}$ of dimension $4\times
4$ is identified as

\begin{equation} (V_{CKM})_{4\times 4} = [(V_{u L}^o\,V_{u L}^{(1)})_{3\times 4}]^T \;(V_{d L}^o\,V_{d L}^{(1)})_{3\times 4}
\end{equation}

\noindent Similar analysis of the couplings of active L-handed neutrinos and L-handed charged leptons to $W$ boson,
leads to the lepton mixing matrix

\begin{equation} ( U_{PMNS} )_{4\times 8}   = [(V_{e L}^o\,V_{e L}^{(1)})_{3\times 4}]^T \;
(U_\nu^o\,U_\nu)_{3\times 8}
\end{equation}

\vspace{5mm}
\section{Conclusions}

We reported an updated and general analysis for the generation of neutrino
masses and mixing within the $SU(3)$ family symmetry model. The
right handed neutrinos $(\nu_e\,\,\nu_\mu\,\,\nu_\tau)_R$, and the vector like completely
sterile neutrinos $N_{L,R}$, the flavon scalar fields and their VEV's introduced to break
the symmetries: $\Phi^u$, $\Phi^d$, $\eta_1$,  $\eta_2$ and $\eta_3$, all together,
yields a $8\times8$ general Majorana neutrino mass matrix with four or five massless neutrinos
at tree level. Therefore, light neutrinos get tiny masses from radiative corrections
mediated by the heavy $SU(3)$ gauge bosons. Neutrino masses and  mixing are extremely
sensitive to the parameter space region, and a global fit for all quark masses and mixing
together with the charged lepton and neutrino masses and lepton mixing is in progress.

\vspace{5mm}
\section*{Acknowledgements}

It is my pleasure to thank the organizers N.S. Mankoc-Borstnik, H.B. Nielsen, M. Y. Khlopov,
and participants for the stimulating Workshop at Bled, Slovenia. This work was
partially supported by the "Instituto Polit\'ecnico Nacional",
(Grants from EDI and COFAA) and "Sistema Nacional de
Investigadores" (SNI) in Mexico.


\end{document}